\documentclass[runningheads]{llncs}
\usepackage[T1]{fontenc}
\usepackage{graphicx}
\usepackage{booktabs}
\usepackage[utf8]{inputenc}

\usepackage{float} 
\usepackage{hyperref}

\begin{document}
\raggedbottom
\title{Profiling Cognitive Offloading in LLM-mediated Synthesis Writing: Volume vs. Content}
\titlerunning{Profiling Cognitive Offloading}
\author{Oleksandra Poquet\inst{1,2}\orcidID{0000-0001-9782-816X} \and
Mani Shankar Nanduri\inst{1}\orcidID{0009-0008-1659-1569} \and
Maria Ximena Salinas Loyer\inst{1}\orcidID{0009-0004-1283-5934}  \and
Matthias Stadler\inst{3}\orcidID{0000-0001-8241-8723}  \and
Michael Sailer\inst{4}\orcidID{0000-0001-6831-5429}\and
Jelena Jovanovic\inst{5,6}\orcidID{0000-0002-1904-0446}}
\authorrunning{Poquet et al.}
\institute{Technical University of Munich, Germany \\
\email{\{sasha.poquet, mani.nanduri, ximena.salinas\}@tum.de}\\ \and
C3L, Adelaide University, Australia\\\and
LMU University Hospital, Germany\\
\email{matthias.stadler@umed.uni-muenchen.de} \and
University of Augsburg, Germany\\
\email{michael.sailer@uni-a.de} \and
University of Belgrade, Serbia\\
\email{jelena.jovanovic@fon.bg.ac.rs} \and
SLATE, University of Bergen, Norway\\}

\maketitle              % typeset the header of the contribution
\begin{abstract}
This study compares two approaches to profiling how learners offload cognitive activity to LLMs during a synthesis writing task. Drawing on Salomon's distributed cognition and the Kintsch and van Dijk model of text comprehension, the study operationalises offloading to an LLM in two ways: as a volume of LLM use and as content of what is offloaded, both along with prior knowledge. Data from 97 university students using a ChatGPT-based interface were analysed using k-means clustering. To capture the content of offloading, their prompts were interpreted as to who performs the activity (active/passive) and at what level of comprehension (local/global). Volume-based profiling (k=4) differentiated learners primarily by prior knowledge, with volume negatively associated with essay authorship. Content-based profiling (k=5) revealed qualitatively distinct patterns of offloading, from vocabulary clarification to active direction of structuring and generation to passive delegation of comprehension at both levels. These patterns reflect different fragmentation of the cognitive process, with significant differences in learning strategies, behavioural markers, and essay authorship. Combining volume and content of offloading could improve future analyses on how LLM use redistributes cognitive activity and its effects on learners.

\keywords{LLM-mediated writing  \and offloading \and distributed cognition}
\end{abstract}

\section{Introduction}

Using large language models (LLMs) in education has a range of ethical, environmental, and privacy implications. When LLMs are used as general-purpose chatbot interfaces without pedagogical restrictions, their effects on human learning are yet to be understood. Identifying and evaluating these effects is a challenge that requires, first, comprehensively describing LLM use through existing theories about learning and cognition, in ways that reflect the novelty of the phenomenon. Motivated by these considerations, this paper analyses the process of LLM use in a learning task in ways that may explain its effects on learners.

Our starting point is that learner use of LLMs changes individual cognitive processes. This change is enabled by cognitive offloading, i.e. the physical alteration of the information processing requirements of a task to reduce cognitive demand \cite{risko}.  Humans offload cognitively from an early age \cite{armitage2020developmental} using body  gestures \cite{goldin2001explaining},  writing \cite{risko2015storing}, other people \cite{wegner1995transactive}, and automated tools \cite{parasuraman1993performance}. Due to this natural propensity, offloading has been well studied. Research predating LLMs established that offloading can impair memory \cite{sparrow2011google} and inflate estimates of what has been retained as own knowledge \cite{fisher2015searching}. Unsurprisingly, similar effects have been corroborated in the studies on LLM use for learning \cite{bastani2024generative,stadler2024cognitive}. 

To guide learner use of LLMs, researchers need to further understand offloading \textit{processes}. Evidence from non-LLM contexts suggests that how offloading processes should be represented is not straightforward. On the one hand, studies in non-LLM contexts show that the \textit{volume} of external tool use affects performance  \cite{grinschgl2023mutual} and memory \cite{sparrow2011google} . On the other hand, studies \textit{also} show that the same volume of offloading can affect what is remembered in different ways, as different mechanisms may be activated \cite{magen2025reduced}. The use of learning strategies when offloading in non-LLM contexts also has differentiated effects, further suggesting that differences in how something is offloaded are important \cite{donet2026effect}.  It is unclear if these two perspectives on offloading reflect different or redundant  information, if \textit{any} use of an LLM or a \textit{particular} use of an LLM is meaningful to capture. 

LLM-focused research analysed both volume of offloading and how learners offload, without directly comparing them. Some studies equate \textit{any} use of the LLMs as indicative of their effect on learning, contrasting LLM use vs no LLM use \cite{stadler2024cognitive,bastani2024generative}. Others characterise the processes of LLM with varied approaches: mapping LLM prompts to the levels in Bloom's taxonomy \cite{furqon2025evaluating},  examining whether learners modify AI output \cite{yang2025modifying}, or inferring learning and meta-cognitive strategies from prompt content \cite{shibani2024untangling,fan2025beware}. No study, as far as we are aware, compared both volume and content of offloading to LLMs, leaving a gap as to how offloading processes are best captured.

Such a comparison is complicated by a theoretical challenge. Existing theories were not designed for tools that afford the range and flexibility of offloading that LLMs enable. Hence, it is unclear how learner-LLM use should be approached analytically. Cognitive load theory (CLT) offers a well-grounded account explaining how types of cognitive load can affect learning \cite{sweller1998cognitive}, also in technology-mediated settings \cite{mayer2022future}. CLT focuses on why a task is cognitively demanding and how to reduce the load through instructional media and tasks, but is less concerned with \textit{how} cognitive activity gets distributed. This limits its utility to describe the process of offloading. Distributed cognition perspectives, on the other hand, are largely socio-cultural. They focus on how cognitive activity gets distributed across people and artefacts \cite{salomon1997distributed}, but pay little attention to individual cognition. This limits their utility in explaining what happens to the individual when offloading takes place. In this paper, we  explore a bridging alternative: a dynamic interactionist view that posits that distributed cognition, individual cognition, and the activity that affords the distribution of a cognitive process \cite{salomon1993no} should all be considered when analysing distributed cognition systems.

Guided by this bridging perspective, we compare insights from examining learner use of LLMs in a synthesis writing task through the volume of offloading vs the content of offloading, while considering individual cognition. Using k-means clustering we profile offloading processes: (1) as number of LLM prompts combined with prior knowledge versus (2) as the content of each prompt, reflecting who performs the cognitive work and at what level of comprehension, combined with prior knowledge. Data for the analysis were collected via a lab-based experiment where 97 university students engaged in a task using a custom-made interface for writing and prompting a general purpose LLM. Profiles identified through each clustering approach were interpreted by comparing learning strategies represented by the prompts, self-reports of learner experience, and behavioural and linguistic markers. The results show that profiling on both volume and content of offloading captures relevant information, suggesting the need to combine them for profiling offloading processes. 

\section{Conceptualising Learner-LLM Use}

To profile offloading during learner use of LLMs, we conceptualise the relationship between learner, LLM, and an AI-mediated activity, drawing on Salomon's ideas about distributed cognition \cite{salomon1993no}. Salomon opposed his contemporaries who privileged socio-cultural contexts when describing how activity gets distributed across tools and among individuals. He argued that the question of where cognition resides "cannot be dealt with in an either (in one's head)/or (distributed) fashion" (p. 111, \textit{ibid.}). His conceptualisation contains three elements of a distributed cognitive system: (1) individual cognition, (2) the activity that affords distribution, and (3) the  distribution of the cognitive process between a learner and a tool. This conceptualisation allows to capture a learner-LLM system. 

First, individual cognition in a learner-LLM system can be considered before and after distributed cognition takes place. A learner enters an activity that can be mediated by an LLM with individual differences that affect their information processing. Then, distributing activity leaves 'cognitive residue'. Salomon proposes that the distribution in the form of the 'division of labour' would lead to helping less skilled individuals access new activities but reduce further development of "the crucial cognitions so off-loaded" (p. 133, \textit{ibid}). Or, the distribution can take a form of a 'shared activity' that may have a more positive effect on learner competency. In the study, we approach learner prior knowledge as a means to represent what an individual brings into the cognitive process. 

Second, the activity that affords the learner to distribute cognition to an LLM needs to be further specified. We propose to treat comprehension as this activity. Comprehension underpins many tasks (writing, programming, problem solving), requires different learning strategies (selection, organisation, and generation of information), and provides a coarse-grained layer applicable to various situations. In our study, learners had to read two texts to integrate them and write an essay, which is an activity that involves meaning making at the level of each text and across them, further externalised through written text. To describe how this activity affords distribution, we draw on the model for text comprehension and production by Kintsch and van Dijk \cite{kintsch1978toward}. Their model explains meaning making at two levels: the micro-structure, where readers resolve local propositions and establish coherence between adjacent sentences, and the macro-structure, where readers construct global meaning through deletion, generalisation, and construction of new propositions across micro-structures. Comprehension can therefore be distributed at either of these levels. 

Finally, the distribution can take different forms depending on the learner's role in the cognitive activity, here at each level of comprehension. The learner may actively participate in some parts of the cognitive activity, for instance, providing their own reasoning while requesting the LLM to elaborate or check it. Such offloaded activity would be shared (i.e. joint), given that the learner maintains parts of the cognitive effort. Alternatively, the learner may be passive, fully removing comprehension activity from their individual cognition, dividing the work rather than sharing it. Introducing the learner's role into the cognitive activity enables capturing Salomon's distinction between shared activity and division of labour. Thus, combining the level of comprehension (micro or macro) with the role of the learner (active or passive) could describe the content of the distributed cognitive process.

\section{Research Question}

The above conceptualisation lends itself to characterising the distributed cognitive process in two distinct ways. As noted earlier, experimental empirical work on offloading in non-LLM contexts suggests that both volume and content of offloading can affect learning. Number of LLM prompts, as the most basic indicator of how much a learner uses the LLM, could capture the volume of offloading and reflect distributed cognition during AI-mediated comprehension. Such a view would be consistent with  a technocentric perspective \cite{salomon2002technology} around the effect of mere tool use on learning as a central theme in the seminal Great Media Debate \cite{clark1994media,kozma1994will}. On the other hand, capturing whether learners actively engage in local or global comprehension, i.e., what parts of comprehension are offloaded, would reflect an alternative view of distributed cognition during AI-mediated comprehension. In this study, we apply both operationalisations to the same data, each combined with prior knowledge, and compare them to understand if they could explain the effect of LLM use on learning. We use clustering to identify offloading profiles based on each approach. This leads to the following research question: 
\paragraph{What profiles of offloading emerge when learner use of LLMs is characterised through (i) the volume versus (ii) what is being offloaded, and do they capture different aspects of the learning experience?} 
    
\section{Methods}
\subsection{Study Description}
Data analysed in this paper were collected during a lab-based study approved by the TUM ethics committee (2024-69-NM-BA).  Participants were university students and received a 20-euro compensation. They were eligible if enrolled in a degree program, proficient in English, and not studying psychology. The latter was due to students' potential familiarity with the instruments administered post-task. The study was facilitated by two researchers in one-hour small-group sessions. 

The final sample included 97 participants, as three students were removed due to incorrect task completion. The students studied Engineering (33.0\%), Business/Economics (20.6\%), Computer Science/Informatics and Social Sciences (both 12.4\%). The sample was almost gender-balanced (55.7\% female and 43.3\% male). The students were 24–26 years old (41.2\%) and 21–23 years old (27.8\%), predominantly at the Master’s level (80.4\%; the remainder being undergraduate). 

Participants received two academic texts with opposing views on the topic of 'Predictive Modelling in Education' and were asked to write a synthesis essay using an interface that embedded access to an LLM. Study facilitators ensured no other tools were used besides our custom-built interface, which included an area to prompt the LLM and a writing pad for typing and pasting. The two texts were designed in line with synthesis-task research \cite{robledo2024strategic}. They were balanced in length and the number of ideas. To ensure readability and difficulty for a university-level audience, both texts were evaluated using the Flesch–Kincaid readability test, with scores within the range typical for upper undergraduate students (Text A: 35.4, Text B: 34.8). The texts are available at https://shorturl.at/s6qKD \href{https://shorturl.at/s6qKD}.

\subsection{Data Analysis: Offloading Profiles}
To profile offloading processes, we applied k-means clustering. First, we clustered the volume of LLM use as the total number of prompts used by the learner in a session along with the total score on prior knowledge test. The 12 users who did not use the LLM were included in the clustering with scores of 0 for the total number of prompts. Second, we clustered the content of LLM use. To do so, each LLM prompt was interpreted using a coding scheme  (Table 1) that represented if learner remained somewhat active or was passive during a comprehension activity, and if the comprehension within a prompt was targetting local or global level. A total of five features were used for this clustering: proportions of prompts (active-local; active-global; passive-local; passive-global) and total score on a prior knowledge test. The 12 participants who did not use the LLM were also included in the clustering with scores of 0 for all prompts.   We compared the clusters across a broad range of measures (Table 2) using Kruskal-Wallis tests with Benjamini-Hochberg correction for multiple comparisons. We opted for this less conservative error correction approach as appropriate for an exploratory study. 

\subsection{Content Analysis of Prompts Used for Clustering }
An experienced researcher conducted content analysis of learner prompts (N=564) to identify how a learner was involved in the comprehension process (actively/ passively) \cite{salomon1993no} and what level of text comprehension the prompt reflected (local/ global)\cite{kintsch1978toward}. Percentages of prompt labels (e.g.,\% of active and local) were used for clustering content-based offloading profiles. Table~\ref{tab:coding} presents the coding framework (\href{https://shorturl.at/s6qKD}{https://shorturl.at/s6qKD}). Reproducibility was assessed via automated labelling via three LLM-based agents (Claude Sonnet 4.6), each given the full coding framework and instructed to label the prompts, check against the framework rules, and revise the annotation, if discrepant. Cohen's kappa was computed between the expert and the majority vote of the three agents. Consistent automated mislabelling was addressed by refining the framework. Further labelling of two batches of prompts  that were not used to refine the framework resulted in an average kappa of 0.64 (passive/global) and 0.61 (local/global). The framework was not further fine-tuned to avoid overfitting.

\begin{table}[H]
\centering
\caption{Coding framework for content-based offloading profiles}
\label{tab:coding}
\footnotesize
\setlength{\tabcolsep}{4pt}
\begin{tabular}{p{1.1cm} p{6.7cm} @{\hspace{0.3cm}} p{3.8cm}}
\toprule
\textbf{Code} & \textbf{Definition} & \textbf{Examples} \\
\midrule
Active & Reflects learning involvement; learner directs the activity: clarifying a knowledge gap, seeking particular type of evaluation or structure,  constraining how text is generated. Coded as active as long as the direction of work is specified by the learner, i.e. some learner effort is present. & ``What does the assumption mean in text A?''; ``I think the main ideas of text B are --- am I right?''; ``Move last paragraph to the beginning.'' \\
\midrule
Passive & Reflects lack of learner involvement; learner does not direct the activity, leaving LLM to perform it without learner-specified constraints. & ``Summarize the following''; ``Write me a 300-word synthesis''; ``Create an outline'' \\
\midrule
Local & Reflects the level of comprehension where activity is conducted. Applies to prompts that include specific information at the micro-level of a word, a paragraph, such as a specific word, sentence, paragraph, concept, or example. & ``How can I improve this phrase?''; ``Summarise the paragraph''; ``What does the text say about the biology students?'' \\
\midrule
Global & Reflects the level of comprehension in relation to the macro-level structure. Applies to prompts that refer to the whole text(s), the essay as a whole, or broad topics spanning both texts. & ``Compare both texts''; ``Write me a 300-word synthesis''; ``Are all main ideas mentioned?'' \\

\bottomrule
\end{tabular}
\end{table}
%\vspace{-8pt}

%A further subset of 47 prompts not used for this discussion  resulted in a kappa of 0.76 for active/passive code and kappa = 0.61 for local/global codes. The remainder full set of non-discussed prompts ($n = 451$), resulted in a kappa of 0.58 for active/passive (85.8\% agreement) and kappa = 0.58 for local/global (80.8\% agreement), indicating moderate agreement. Disagreements arose primarily on short, context-poor prompts that were not parsed by automated agents. The framework was not fine-tuned further to avoid being overly specific to the specific domain in the study.

\subsection{Measures to describe offloading profiles}
To interpret and compare the clusters, we have used self-reported, behavioural, linguistic, and content-based measures, presented in Table ~\ref{tab:measures}. Self-reported measures included a prior knowledge pre-test on the topic of predictive modelling and a post-test NASA Task Load Index \cite{hart1986nasa}, capturing mental demand, temporal demand, performance, effort, and frustration during the task (internal consistency ($\alpha$ = .68)). 

\begin{center}
\footnotesize
\refstepcounter{table}
\textbf{Table \thetable.} Measures for describing offloading profiles
\label{tab:measures}
%\vspace{4pt}
\centering
\small
\begin{tabular}{p{2.5cm}p{7cm}@{\hspace{0.3cm}}r  @{\hspace{0.3cm}}r  @{\hspace{0.2cm}}r}
\toprule
\textbf{Measure} & \textbf{Description} & \multicolumn{3}{c}{\textbf{Descriptives}} \\
\midrule
\multicolumn{5}{l}{\textbf{Self-reported measures}} \\
\midrule
Prior knowledge & AILIT-S \cite{hornberger2025development}; factual questions about AI and modelling & \multicolumn{3}{p{2.5cm}}{\textit{M}=0.6, \textit{SD}=0.2} \\
\midrule
Cognitive load & NASA TLX \cite{hart1986nasa}; mental demand, temporal demand, performance, effort, frustration ($\alpha$=.68) & \multicolumn{3}{p{2.5cm}}{\textit{M} range 2.5--3.3} \\
\midrule
\multicolumn{5}{l}{\textbf{Behavioural and linguistic measures}} \\
\midrule
Prompts & Total prompts during the task & \multicolumn{3}{p{2.5cm}}{\textit{M}=5.8, \textit{SD}=6.7} \\
\midrule
Time on task & Total time spent on the task, minutes & \multicolumn{3}{p{2.5cm}}{\textit{M}=29, \textit{SD}=9} \\
\midrule
Prompt pacing & Harmonic mean of inter-prompt intervals, minutes & \multicolumn{3}{p{2.5cm}}{\textit{Mdn}=1.5, \textit{IQR}=0.7-8} \\
\midrule
Text-to-prompt borrowing & Proportion of prompt words in unique lexical overlaps (2--5 n-grams) with source texts \cite{keck2006use} & \multicolumn{3}{p{2.5cm}}{\textit{M}=0.35, \textit{SD}=0.26} \\
\midrule
LLM-to-essay borrowing & Proportion of student-authored sentences; If $>$50\% of text snippets has 5-gram overlap with LLM the text is labelled as LLM-derived. & \multicolumn{3}{p{2.5cm}}{\textit{M}=66.3\%, \textit{SD}=42.1\%} \\
\midrule
Connective density & Discourse connectives per sentence, capturing use of integrative words, applied to the student produced portions of the text & \multicolumn{3}{p{2.5cm}}{\textit{M}=0.31, \textit{SD}=0.19} \\
\midrule
\multicolumn{2}{l}{\textbf{Learning strategies (per prompt)}} & \textbf{Total} & \textbf{M} & \textbf{SD} \\
\midrule
Clarify & \multicolumn{1}{p{6.2cm}}{Asks to clarify vocabulary or factual gaps} & 52 & 0.54 & 1.00 \\
Select & \multicolumn{1}{p{6.2cm}}{Seeks key points or information in the source} & 30 & 0.31 & 0.89 \\
Summarise & \multicolumn{1}{p{6.2cm}}{Asks to compress text into a shorter version} & 57 & 0.59 & 1.84 \\
Comprehend & \multicolumn{1}{p{6.2cm}}{Asks for facts explained in the source text} & 67 & 0.69 & 1.72 \\
Structure & \multicolumn{1}{p{6.2cm}}{Seeks to organise text or create connections} & 51 & 0.53 & 1.33 \\
Generate & \multicolumn{1}{p{6.2cm}}{Asks to generate text} & 132 & 1.36 & 2.10 \\
Evaluate & \multicolumn{1}{p{6.2cm}}{Asks to evaluate pasted text} & 40 & 0.41 & 1.00 \\
Paste & \multicolumn{1}{p{6.2cm}}{Pastes text without instruction} & 61 & 0.63 & 1.20 \\
Non-op & \multicolumn{1}{p{6.2cm}}{Asks about word count or grammar} & 74 & 0.76 & 1.75 \\
\bottomrule
\end{tabular}
\end{center}

Behavioural measures included the number of prompts, total time on task, and individual prompt pacing. Linguistic measures included borrowing from the source text to prompt,  borrowing from LLM responses to the essay, and connective density. Connective density, calculated on student-written essay segments, represents independent writing through the use of discourse markers that signal integrative relationships between ideas (e.g., "however," "therefore," "in contrast," "as a result"). It was selected from a range of linguistic measures applied to the student-written parts of the essay, identified as not being copied verbatim. 

Profiles were also compared around the learning strategies applied by each learner to better understand what learners were doing. These categories offer additional qualitative descriptions and are not intended as theoretical insight. Each prompt was assigned to one of nine mutually exclusive categories to reflect the learning strategy used to advance the writing task (Table~\ref{tab:measures}). Among them, seven categories map to learning strategies grounded in Mayer's \cite{mayer1996learners} selecting -organising - integrating (SOI) framework. Two additional categories capture prompts that do not involve a learning strategy. The framework was developed iteratively by an experienced researcher, who initially labelled prompts thematically and then mapped emerging themes to the SOI framework. Following the process described in Section 4.3, the expert and LLM-based agents reached Cohen's kappa 0.68 for 168 prompts ( \href{https://shorturl.at/s6qKD}{https://shorturl.at/s6qKD}).

\subsection{Limitations}
There are several important limitations of the study. The sample and cluster sizes limit statistical power, particularly for detecting differences in self-reported measures. Cluster validation has not been done using learning outcomes since modelling them properly requires substantially expanding the paper scope. Despite our attempt to represent LLM-mediated activity in a generalisable manner, other tasks may require other lenses to represent offloading patterns. 

\section{Results}

\subsection{Volume-based Offloading Profiles}

Volume-based offloading profiles were represented by four clusters (Table~\ref{tab:solA}). The number of clusters was selected based on the elbow method, silhouette scores, and cluster interpretation. Within cluster sum of squares stabilised after k=3–4, with silhouette scores 0.481 for three clusters and 0.412 for four clusters. Four-cluster solution was chosen due to a better differentiation within a larger cluster. 

The clusters significantly differed both in prior knowledge of the learners whose offloading processes they represented, as well as their volume of LLM use. Hence, they were interpreted as representative of learner-LLM systems with \textit{High Prior Knowledge-Minimal Offloading} (n=51), \textit{High Prior Knowledge-Moderate Offloading} (n=17), \textit{Low Prior Knowledge - Low Offloading} (n=22), and \textit{Heavy Offloading} (n=7). The use of LLM across volume-based profiles varied, from 2 prompts in the low offloading group to 25 prompts in the heavy offloading group. 

The volume of offloading related to the essay authorship. The more LLM was used, the lower was student authorship score. This indicates that more offloading meant fewer instances of own knowledge construction. Volume of offloading was also closely related with the prompt pacing. Higher offloading meant lower intervals between LLM use, with the entire writing process done entirely through offloading. For instance, minimal offloading was punctured by an average of 5 minutes between moments of offloading, whereas moderate offloading had a faster pace of less than a minute between LLM use on average. The profiles were not significantly different on the number of prompts for local or global level of comprehension, or the number of prompts signalling active role of a learner. Differences in learning strategies between the clusters were non-informative.

Connective density also varied significantly across profiles. This was negatively correlated with prior knowledge and not the number of prompts (\textit{r} = -.30, \textit{p} = 0.007). This is counterintuitive; potentially, more coherent discourse required fewer connectives. Low connective density of heavy offloaders should be interpreted with caution since they contributed 2\% of the essay. 

\begin{table}[H]
\caption{Volume-based offloading profiles; means of descriptive measures}
%Significance: BH-corrected Kruskal--Wallis, * \textit{p}~<~.05, ** \textit{p}~<~.01, *** \textit{p}~<~.001.
\label{tab:solA}
\centering
\footnotesize
\resizebox{\columnwidth}{!}{%
\begin{tabular}{p{3.2cm} r r r r c}
\toprule
 & \textbf{High knowledge} & \textbf{Low knowledge} & \textbf{High knowledge} & \textbf{Heavy} & \\
 & \textbf{minimal offloading} & \textbf{low offloading} & \textbf{moderate offloading} & \textbf{offloading} & \\
 & \textit{n}=51 & \textit{n}=22 & \textit{n}=17 & \textit{n}=7 & \textit{p} \\
\midrule
\multicolumn{6}{l}{\textbf{Clustering variables}} \\[2pt]
Prior Knowledge & .75 & .34 & .73 & .51 & *** \\
No.\ prompts & 2.4 & 4.3 & 10.1 & 25.1 & *** \\[4pt]
\multicolumn{6}{l}{\textbf{Learning strategies (\% of prompts, LLM users only)}} \\[2pt]
Clarify & 21 & 9 & 7 & 3 & \\
Comprehend & 14 & 22 & 5 & 12 & \\
Select & 7 & 5 & 7 & 3 & \\
Summarize & 7 & 14 & 8 & 12 & \\
Structure & 2 & 10 & 13 & 8 & *** \\
Generate & 17 & 16 & 30 & 25 & *** \\
Evaluate & 10 & 4 & 9 & 5 & \\
Paste & 11 & 17 & 10 & 8 & \\
Non-op & 12 & 4 & 11 & 21 & *** \\[4pt]
\multicolumn{6}{l}{\textbf{Cognitive activity (\% of prompts, LLM users only)}} \\[2pt]
Passive & 77 & 84 & 72 & 82 & \\
Global & 68 & 74 & 66 & 72 & \\[4pt]
\multicolumn{6}{l}{\textbf{Behavioural traces}} \\[2pt]
Essay authorship (\%) & 80 & 81 & 30 & 2 & *** \\
Prompt pacing (min) & 5.00 & 3.76 & 1.16 & 0.53 & *** \\
Text-to-prompt & 37 & 28 & 40 & 35 & \\
Connective density & 1.51 & 2.68 & 2.10 & 1.00 & * \\[4pt]
\multicolumn{6}{l}{\textbf{Self-reported learner experience}} \\[2pt]
Mental demand & 3.18 & 3.41 & 3.41 & 3.57 & \\
Time pressure & 3.41 & 3.18 & 3.12 & 3.57 & \\
Effort & 3.14 & 3.09 & 3.35 & 3.14 & \\
Frustration & 2.29 & 2.91 & 2.29 & 2.57 & \\
Perceived success & 3.16 & 3.09 & 3.53 & 2.86 & \\
\bottomrule
\end{tabular}
}%
\end{table}

Self-reported cognitive load showed no significant differences. There are insufficient observations in each cluster to capture differences, but it is worth noting that high offloading participants reported the highest mental demand, compared with learners who had not offloaded or who had low prior knowledge. This suggests that managing high-paced offloading may still be perceived as demanding.

\subsection{Content-based Offloading Profiles}

Content-based profiling resulted in five clusters (Table~\ref{tab:solB}). Within-cluster sum of squares flattened after k=4–5 (26.1\% and 22.1\% reductions), with silhouette scores of 0.376 for four clusters and 0.424 for five clusters. Five-cluster solution grouped all learners who completed the task without LLM as a distinct profile. 

\begin{table}[H]
\caption{Content-based offloading profiles; means of descriptive measures}
%Significance: BH-corrected Kruskal--Wallis, * \textit{p}~<~.05, ** \textit{p}~<~.01, *** \textit{p}~<~.001, \dag~\textit{p}~<~.10. --- = not applicable (non-users or zero prompts).
\label{tab:solB}
\vspace{-4pt}
\centering
\footnotesize
\resizebox{\columnwidth}{!}{%
\begin{tabular}{p{3.0cm} r r r r r c}
\toprule
 & \textbf{Full} & \textbf{Comprehension} & \textbf{Strategic} & \textbf{Clari-} & \textbf{No-}& \\
 & \textbf{Offloading}& \textbf{Offloading}& \textbf{Offloading}& \textbf{fying}& \textbf{offloading}& \\
 & \textit{n}=41 & \textit{n}=15 & \textit{n}=12 & \textit{n}=17 & \textit{n}=12 & \textit{p} \\
\midrule
\multicolumn{7}{l}{\textbf{Clustering variables}} \\[2pt]
Prior knowledge& .66 & .60 & .58 & .66 & .63 & \\
Active, \% prompts  & 4 & 12 & 36 & 81 & --- & *** \\
Local, \% prompts & 7 & 60 & 22 & 85 & --- & *** \\[4pt]
\multicolumn{7}{l}{\textbf{Learning strategies (\% of prompts, LLM users only)}} \\[2pt]
Clarify & 2 & 5 & 3 & 71 & --- & *** \\
Comprehend & 16 & 12 & 6 & 10 & --- & \\
Select & 6 & 2 & 8 & 4 & --- & \\
Summarize & 7 & 27 & 4 & 0 & --- & ** \\
Structure & 6 & 5 & 19 & 2 & --- & *** \\
Generate & 27 & 23 & 24 & 8 & --- & *** \\
Evaluate & 7 & 5 & 13 & 0 & --- & ** \\
Paste & 15 & 15 & 5 & 0 & --- & *** \\
Non-op & 15 & 6 & 19 & 4 & --- & * \\[4pt]
\multicolumn{7}{l}{\textbf{Behavioural traces}} \\[2pt]
No.\ prompts & 5.5 & 8.9 & 12.9 & 2.9 & 0 & *** \\
Essay authorship (\%) & 61 & 58 & 24 & 91 & 100 & *** \\
Prompt pacing (min) & 2.16 & 4.75 & 1.05 & 5.18 & --- & *** \\
Text-to-prompt (\%)& 41 & 45 & 32 & 38 & 0 & *** \\
Connective density & 1.87 & 2.33 & 2.18 & 1.56 & 1.64 & \\[4pt]
\multicolumn{7}{l}{\textbf{Self-reported learner experience}} \\[2pt]
Mental demand & 3.41 & 3.47 & 3.00 & 3.24 & 3.08 & \\
Time pressure & 3.44 & 3.53 & 2.92 & 3.29 & 3.08 & \\
Effort & 3.29 & 3.20 & 2.83 & 3.18 & 3.00 & \\
Frustration & 2.49 & 2.47 & 2.50 & 2.41 & 2.33 & \\
Perceived success & 3.22 & 3.13 & 3.33 & 3.24 & 2.92 & \\[4pt]
\bottomrule
\end{tabular}
}%
\end{table}

Profiles did not differ in prior knowledge, despite it being a clustering variable. The proportions of what was offloaded dominated cluster differences. Notably, even though the number of prompts was not used for clustering, some content-based offloading profiles were differentiated by the volume of LLM use. 

The five profiles represented qualitatively distinct patterns of how cognitive activity was distributed, with significant differences in learning strategies. We interpret the offloading processes in these learner-LLM systems as follows. \textit{Full Offloading} (n=41) had learners passively involved (passive prompts at 96\%) using LLMs to hone on the comprehension at the global level (93\%). That is, they delegated comprehension, generation, and heavily pasted the text back-and-forth. \textit{Clarifying Offloading} processes (n=17) had learners applying some effort (active prompts at 81\%) mostly at the local level of comprehension (85\%), using the LLM almost exclusively for vocabulary and concept clarification. \textit{Comprehension Offloading} processes (n=15) were characteristic of passively involved learners (passive prompts at 88\%) both at micro- and macro- levels of comprehension (60\% local). They relied heavily on summarisation and requests for specific information, instead of locating it themselves. \textit{Strategic Offloading} processes (n=12) had the highest proportion of offloading at the global comprehension level but with active learner participation  (36\% active). These learners directed structure of the essay and generation at high volume and fast pace. \textit{Non-users} (n=12) completed the task independently.

Some content-based profiles have significant differences in the prompt numbers. Strategic Offloaders offloaded at the highest volume (\textit{M}=12.9) and the fastest pace (\textit{M}=1.05 min), followed by the Full Offloaders who prompted moderately (M=5.5). These two groups have similarly high levels of prior knowledge but they offloaded very differently, qualitatively speaking. Strategic offloaders have a high level of learner engagement in directing the specifics of the comprehension activity: despite their high level of offloading, they are likely not completely removing comprehension, maintaining some of it and distributing the rest. Full offloaders do not at all direct, their active level is at 4\% of all prompts on average, and they do not engage in offloading comprehension at the micro-level (7\% only). These processes are contrasted by the Clarifiers who essentially use LLMs to reduce intrinsic cognitive load when accessing complex or unknown ideas, conducted through low-level of LLM prompting (Mprompts=2.9). None of these qualitative  differences are conveyed by volume-based profiling.

Self-reported cognitive load showed no differences across content-based profiles. As with volume-based profiling, cluster sizes limited statistical power. Descriptively speaking, learners who employed Strategic Offloading reported the lowest mental demand (M=3.00), time pressure (M=2.92), and effort (M=2.83), but the highest frustration. Overall, the trends might indicate that perceptions of mental demand, pressure, effort, and frustration are driven by various needs for information processing. Hence, new self-reported instruments are needed to understand the experience of offloading dynamically.

Content-based offloading profiles also differed on essay authorship, prompt pacing, text-to-prompt borrowing, and six of nine learning strategies (all \textit{p}~<~.01). For instance, strategic offloaders wrote a quarter of the essay independently, whereas both full and comprehension offloaders produced around 60\% of the text themselves. Connective density was similarly not differentiated across profiles, and considering its correlation with prior knowledge, it may be assumed that content-based profiles subsume differences stemming from prior knowledge. 

\subsection{Two Approaches: Capturing Fragmentation}

\begin{figure}
    \centering
    \includegraphics[width=1\linewidth]{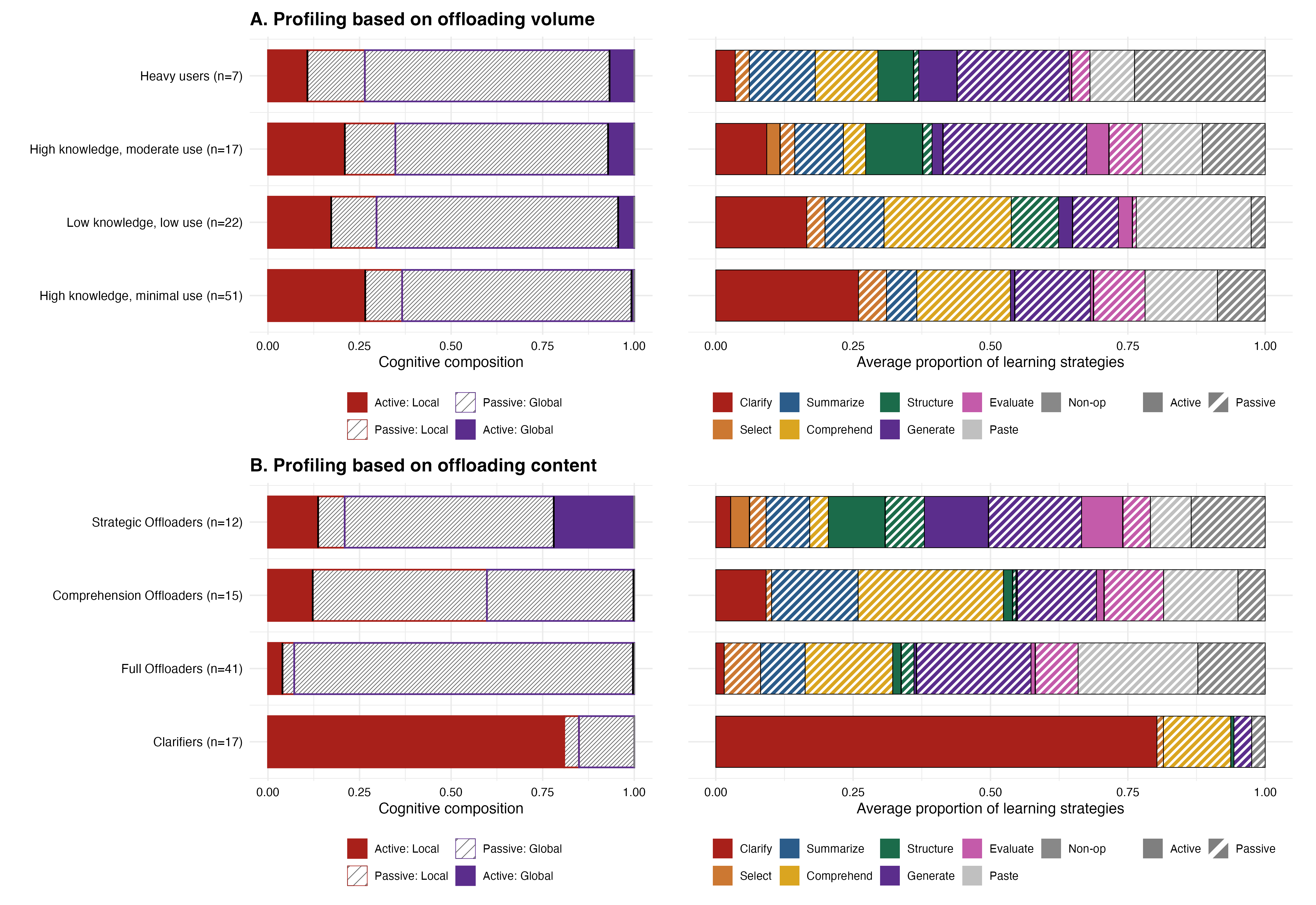}
   \caption{Fragmentation in Offloading, based on volume (A, top) and content (B, bottom) of LLM use. Stripes capture parts where the learner was passive,Solid colours represent that the learner was active. Profiles are ordered from the lowest to the highest prompts}
    \label{fig:offloading_profiles}
\end{figure}

To compare approaches to profiling offloading, Figure ~\ref{fig:offloading_profiles}, plots them side by side. Panel A captures volume-based offloading profiles, ordered from low LLM use at the bottom for \textit{High Knowledge, minimal use} offloading to the \textit{Heavy LLM use} at the top. Panel B similarly presents content-based offloading profiles ordered from the lowest LLM use for \textit{Clarifying offloading} processes at the bottom and highest LLM use at the top for \textit{Strategic Offloading}. We use stripes to visualise where learner effort is minimal (passive prompt labels), demonstrating where individual cognitive process  fragments when cognitive activity is distributed to LLMs. Active prompt labels maintain their solid colour. 

\vspace{-4pt}
\vspace{-17pt}
 
These visualisations demonstrate that volume of offloading masks differences that are visible when profiling what is offload rather than how much.  For instance, Panel~B shows how \textit{Strategic Offloaders} actively structure information and provide specific directives for text generation, while selectively delegating other operations to the tool. This pattern contrasts sharply with \textit{Full offloading} and \textit{Comprehension offloading}, where learners delegate entire segments of the comprehension and production process and participate actively in few, if any, of the task's cognitive operations.

\section{Discussion and Implications}
Motivated by competing evidence from the offloading literature as to whether offloading process should be operationalised via its volume or what is offloaded, we analysed LLM use by learners in a writing synthesis task. We use the same data to profile their offloading with these two approaches. We adopted Salomon's ideas about distributed cognition \cite{salomon1993no} to analytically represent learner-LLM distributed cognitive system as individual cognition via prior knowledge, the activity that affords distribution, and the distribution of the cognitive process. Volume-based profiling combined prior knowledge with the amount of LLM use. Content-based profiling combined prior knowledge with variables assigned to the prompt level, drawn from Kintsch and van Dijk's levels of comprehension (local or global) and the role of a learner in cognitive activity (active or passive). 

The two operationalisations did not capture redundant information. Volume-based profiles appear to have differentiated aspects of offloading associated with prior knowledge rather than offloading volume. In our data, offloading volume is negatively associated with student essay authorship, indicating the decrease of explicit knowledge construction when working on the final artefact. Content-based profiles differentiated the process by capturing how the comprehension process was fragmented by different offloading strategies. Although volume was not a clustering variable in the content-based solution, some of the resulting profiles differed in prompting volume, suggesting that sometimes what is offloaded constrains how much is offloaded. We also show that qualitative differences exist within similar volumes of offloading. Strategic offloading, involving iterative structuring and generation from an actively engaged learner, required higher offloading volume. So did Full offloading, but without engaging the learner in the comprehension process.

These patterns, captured through fragmentation, can be related to Salomon's \cite{salomon1993no} distinction between division of labour, where the cognitive process is handed off, and shared activity, where it is jointly maintained. While the fragmentation view suggests that the volume of LLM use could often still be a proxy for what is likely ineffective offloading, our findings highlight the relevance of the content aspect. The parts of the cognitive process that are retained by the individual are important to know when evaluating high or moderate use of LLM. Low volume of LLM use is most likely to reflect shared activity, but high volume can accompany either division of labour or shared work, depending on the patterns in fragmentation. In either case, treating LLM use as binary, through a techno-centric perspective, does not offer insight that differentiates variations in offloading processes.

From a psychological perspective, the distinction between volume-based and content-based profiles poses further questions. One assumption may be that offloading reduces cognitive demand \cite{risko}. From a CLT perspective, reducing task demands does not necessarily mean that the cognitive resources freed through offloading will be used for the processes needed for comprehension and schema construction \cite{sweller1998cognitive}. Moreover, offloading dynamically may introduce other unexpected cognitive demands that still require cognitive resources from the learner. Understanding the cause of this potential effect and minimising it may then be further needed. In either case, fragmentation could be approached as a proximal outcome of cognitive offloading, potentially helpful in explaining how different offloading strategies redistribute the cognitive work involved in comprehension, as well as subsequent choices and behaviour. 

For educational practice, these findings suggest that interventions targeting the volume of LLM use, such as limiting the number of prompts, may be insufficient, since the same volume could describe rather different offloading processes. More targeted approaches would attend to what learners delegate: whether they retain active involvement in structuring and evaluating, or hand off entire segments of the comprehension process. Tools that make the distribution of cognitive work visible to learners and instructors could potentially support this, though more work is needed to evaluate offloading processes and their differentiated effects on learners. Future work should examine whether fragmentation patterns predict learning outcomes directly, and whether they hold across tasks with different cognitive demands.

\bibliographystyle{splncs04}
\bibliography{mybib}
\end{document}